\documentclass[twocolumn,preprintnumbers,amsmath,amssymb]{revtex4}
\usepackage{amsmath,amssymb,graphics,epsfig,subfigure}
\usepackage{cases}
\usepackage{color}
\usepackage[colorlinks,
            linkcolor=blue,
            anchorcolor=blue,
            citecolor=green
            ]{hyperref}
\usepackage[colorlinks,linkcolor=blue,anchorcolor=blue,citecolor=green]{hyperref}

\begin{document}
\renewcommand{\baselinestretch}{1.3}

\title{Constraint on the radius of five-dimensional dS spacetime with GW170817 and GRB 170817A}

\author{Zi-Chao Lin$^a$, 
        Hao Yu$^a$, 
        Yu-Xiao Liu$^a$$^b$\footnote{liuyx@lzu.edu.cn, corresponding author}}

\affiliation{$^{a}$Institute of Theoretical Physics $\&$ Research Center of Gravitation, Lanzhou University, Lanzhou 730000, China\\
$^{b}$Key Laboratory for Magnetism and Magnetic of the Ministry of Education, Lanzhou University, Lanzhou 730000, China}

\begin{abstract}
The recent detections of the gravitational wave (GW) event GW170817 and its electromagnetic counterpart GRB 170817A produced by a binary neutron star (NS) merger is a new milestone of multimessenger astronomy. The time interval between these two signals has attracted widespread attention from physicists. In the braneworld scenario, GWs could propagate through the bulk while electromagnetic waves (EMWs) are bounded on the brane, i.e., our Universe. Therefore, the trajectories of GWs and EMWs may follow different pathes. If GWs and EMWs are originated simultaneously from the same source on the brane, they are expected to arrive at the observer successively. Consequently, the time delay between GW170817 and GRB 170817A may carry the information of the extra dimension. In this paper, we try to investigate the phenomenon in the context of a five-dimensional dS ($\text{dS}_5$) spacetime. We first study two special Universe models, i.e., de Sitter and Einstein-de Sitter models, and calculate the gravitation horizon radius in each case. For the real Universe, we then consider the $\Lambda$CDM model. Our results show that for the de Sitter model of the Universe, the $\text{dS}_5$ radius could not contribute to the time delay. With the data of the observation, we constrain the $\text{dS}_5$ radius to $\ell\gtrsim7.5\times10^{2}\,\text{Tpc}$ for the Einstein-de Sitter model and $\ell\gtrsim2.4\times10^{3}\,\text{Tpc}$ for the $\Lambda$CDM model. After considering the uncertainty in the source redshift and the time-lags given by different astrophysical processes of the binary NS merger, we find that our constraints are not sensitive to the redshift in the range of ($0.005$, $0.01$) and the time-lag in the range of ($-100$\,s, $1.734$\,s).
\end{abstract}



\maketitle

\section{Introduction}

The nature of gravitational waves (GWs) is the perturbations of spacetime. But such perturbations are so weak that it took humans decades to detect them. Since GWs were first detected by the LIGO and Virgo collaborations in 2015~\cite{Abbott1}, in just two or three years, they have detected more than ten GW events through gravitational observation. These events involve the binary black hole mergers as well as the binary neutron star (NS) merger~\cite{Abbott1,Abbott2,Abbott3,Abbott4,Abbott5,Abbott6}. The latter is a special case, because the binary NS is widely expected to radiate both short gamma ray bursts (SGRBs) and GWs during its merger~\cite{Paczynski1,Eichler1,Narayan1,Barthelmy1,Shibata1,
Rezzolla1,Ciolfi1,Tsang1,Rezzolla2,Paschalidis1}. In the event GW170817, the collaboration of the LIGO-Virgo detectors, the {\em Fermi} Gamma-ray Burst Monitor (GBM), and the spectrometer on board {\em INTEGRAL} Anti-Coincidence Shield (SPI-ACS) found that the gravitational signal GW170817 originated from the merger of a binary NS is followed by the electromagnetic (EM) signal GRB 170817A~\cite{Abbott6,Coulter1,Pan1,Abbott7}. Since the source of GRB 170817A is very close to the source of GW170817 and the time interval between the two signals is only $1.7$ seconds, most people believe that GRB 170817A is an EM counterpart of GW170817~\cite{Abbott6,Coulter1,Pan1,Abbott7,Abbott8}.

Although it is now widely accepted that a binary NS merger could emit GWs and SGRBs, the time interval between the events GW170817 and GRB 170817A is still nerve-racking. So far, there are plenty of researches attempting to explain the time delay. From the perspective of the gravity itself, the linearized Einstein equation has various forms in different modified gravities, so the propagation speed of the introduced extra GW polarization modes could deviate from the speed of light~\cite{Comelli1,Will1,Saltas1,Chen1,Rizwana1,Andriot1, Ghosh1}. On the other hand, the binary NS could undergo some exotic astrophysical processes during its merger. In these processes, the generation of the GWs and SGRBs may not be simultaneous~\cite{Shibata1,Rezzolla1,Paschalidis1,Rezzolla2,Ciolfi1,Tsang1}.

In higher-dimensional theories, the time delay between the GWs and SGRBs is also reasonable~\cite{Chung1,Caldwell1,Wang2002,Yu1,Visinelli1,Ishihara1}. After one hundred years of development of the extra dimensional theories from Kaluza-Klein (KK) theory~\cite{Kaluza1,Klein1,Klein2}, most famous higher-dimensional theories agree with that our Universe is a four-dimensional hypersurface (called brane) embedded in a higher-dimensional spacetime. In this scenario, the elementary particles and interactions in the Standard Model are confined on the hypersurface, while the gravity could propagate through the bulk. It indicates the possibility that the trajectory of five-dimensional null geodesics might deviate from light, which hence may result in the time delay~\cite{Caldwell1,Yu1,Visinelli1}.

Apart from the time delay, extra dimensions have another important effect on GWs, which may reveal the number of spacetime dimensions. In higher-dimensional theories, since GWs could propagate through the bulk, they will leak into extra dimensions (called gravitational leakage) when propagating in our Universe. Therefore, for a given higher-dimensional theory, the amplitude of GWs will decay faster than the expectation in the four-dimensional theory. It is then expected that the gravitational leakage will reduce the amplitude of the observed GWs and make the four-dimensional observer to misjudge the travel distance of the GWs. Generally speaking, the more extra dimensions in the spacetime, the faster the amplitude of GWs will decay during their propagation. According to this and the data of GW170817, the modified amplitude of the GWs in a specific higher-dimensional theory was used to constraint the number of spacetime dimensions~\cite{Pardo1,Abbott10,Dvali1,Deffayet1}. The results shown that, the number of spacetime dimensions could be larger than four.

On the other hand, it is well known that some extra dimensional theories could unify gravity and electromagnetism~\cite{Kaluza1,Klein1,Klein2} and solve the huge hierarchy between the fundamental scales of gravity and electromagnetism~\cite{ArkaniHamed1998rs,Randall1,Randall2,Guo2018}. All of these features indicate that it is important to investigate the structure of extra dimensions and it is worth paying close attention to revealing the information of it from the time delay and its other effects.

Recently, a five-dimensional theory with static spherically symmetric anti--de Sitter (AdS) spacetime was studied in Refs.~\cite{Caldwell1,Yu1,Visinelli1}. The brane is embedded as a curved hypersurface in this model and the null trajectories of GWs and electromagnetic waves (EMWs) are concerned. Similar to most extra-dimensional theories, EMWs are confined on the brane while GWs could propagate through the bulk. It was found that the gravitational horizon radius on the brane is effected by the structure of the bulk spacetime. Then the discrepancy between the gravitational horizon radius and photon horizon radius, which may eventually result in the time delay, could be used to reveal the feature of the $\text{AdS}_{5}$ radius. In Ref.~\cite{Yu1}, the authors analyzed such model with the assumption that our Universe is closed. The cases that our Universe is dominated by either dark energy or nonrelativistic matter were discussed. In both cases, the present-day spatial curvature density is constrained to the scale of $10^{-10}$ which is much smaller than the result obtained by the {\em Planck} collaboration~\cite{Planck1,Planck2}. Later, it was found that the time delay also occurs even if our Universe is flat~\cite{Visinelli1}. In this case, the $\text{AdS}_{5}$ radius is required to be smaller than $0.535\,$Mpc at $68\%$ confidence level.

In this work, we focus on the braneworld embedded in a five-dimensional de Sitter ($\text{dS}_{5}$) spacetime, which is on account of the following motivations. The cosmological observations indicate that the phase of our very early Universe is a dS phase and that the Universe may ultimately dominated by the dark energy (which is also a dS phase). So it is worth generalizing the bulk spacetime into a $\text{dS}_{5}$ case where the braneworld could be realized by a quantum creation from the holographic dS/CFT correspondence~\cite{Hull1,Strominger1}. On the other hand, it was already found that the $\text{dS}_{4}$ brane with a modified Friedmann-Lema\^{\i}tre-Robertson-Walker (FLRW) equation could be constructed in the $\text{dS}_{5}$ bulk~\cite{Verlinde1,Nojiri1,Nojiri2,Addazi1}. The mass, entropy, holography, and other properties of the five-dimensional Schwarzschild-dS black hole were also discussed in Refs.~\cite{Bousso1,Danielsson1,Balasubramanian1,Cai1,Addazi1}. All these breakthroughs indicate that studying the structure of the $\text{dS}_{5}$ spacetime has a guiding significance for us to understand our Universe.

Inspired by Refs.~\cite{Caldwell1,Wang2002,Yu1,Visinelli1}, we then try to constrain the $\text{dS}_{5}$ radius with the observed time delay between the detections of GW170817 and GRB 170817A. For a $\text{dS}_{5}$ bulk, we would like to embed our Universe inside the cosmological horizon so that the scale factor of our Universe could increase from $a(t_{0}=0)=0$. Indeed, it might require a large $\text{dS}_{5}$ radius. On the other hand, it has been proved that the standard cosmology model will be recovered on the brane as long as the $\text{dS}_{5}$ radius is extremely large~\cite{Verlinde1,Nojiri1,Nojiri2,Addazi1}. Therefore, if such braneworld is true, the constraint from the observed time delay should allow the existence of the large $\text{dS}_{5}$ radius. We adopt two special toy models and $\Lambda$CDM model for our Universe to calculate the discrepancy between the gravitational horizon radius and photon horizon radius. With the event GW170817/GRB 170817A, we respectively obtain two lower bounds on the $\text{dS}_{5}$ radius, $\ell\gtrsim7.5\times10^{2}\,\text{Tpc}$ (Einstein-de Sitter model) and $\ell\gtrsim2.4\times10^{3}\,\text{Tpc}$ ($\Lambda$CDM model), which are consistent with the above analysis.

The following context is arranged as follows. In Sec.~\ref{sec2}, we construct the brane model in a five-dimensional static spherically symmetric dS spacetime, where our Universe is embedded. We then give the abstract forms of the gravitational horizon radius and photon horizon radius in Sec.~\ref{sec3}. In order to express the gravitational horizon radius in a practical form, we convert the unknown quantity into the observable quantities in Sec.~\ref{sec4}. Combined with the data of GW170817 and GRB 170817A, we give the constraint on the $\text{dS}_{5}$ radius in Sec.~\ref{sec5}. Finally, we make a short conclusion in Sec.~\ref{sec6}.

\section{Embedded Universe}\label{sec2}

We start from a five-dimensional spherically symmetric spacetime with the metric:
\begin{eqnarray}\label{ds0}
ds^{2}_{5}=-f(R)dT^{2}+f(R)^{-1}dR^{2}+R^{2}d\Sigma_{k}^{2},
\end{eqnarray}
where $T$ is the coordinate used to denote the sequence of events, $d\Sigma_{k}^2$ is a metric on a locally homogeneous three-dimensional surface of constant curvature $k$:
\begin{eqnarray}
d\Sigma^{2}_{k}=\frac{1}{1-kr^{2}}dr^{2}+r^{2}d\theta^{2}+r^{2}\text{sin}^{2}\theta d\phi^{2},
\end{eqnarray}
and $R$ is the spatial coordinate denoting the radial coordinate distance of the hypersurface $\Sigma_{k}$ from the coordinate origin. Here and after, we assume that the geometry of the bulk is dominated by a bulk cosmological constant. Then we can get a series of Schwarzschild-like solutions of the metric by solving the Einstein equation with the bulk cosmological constant:
\begin{equation}
f(R)=\left\{
\begin{aligned}
&k~(k\neq0) &\text{for}&\text{ Minkovski bulk}\\
&k-\frac{R^{2}}{\ell^{2}}-\frac{\mu}{R^{2}} &\text{for}&\text{ dS bulk}\\
&k+\frac{R^{2}}{\ell^{2}}-\frac{\mu}{R^{2}} &\text{for}&\text{ AdS bulk}
\end{aligned}
\right..\label{ds1}
\end{equation}
where the parameter $\ell$ is the $\text{dS}_{5}$ ($\text{AdS}_{5}$) radius and $\mu$ is the Schwarzschild-like mass. For a $\text{dS}_{5}$ spacetime, there are a cosmological horizon at
\begin{equation}
R_{\text{ch}}=\sqrt{\left(k\ell^{2}+\sqrt{k^{2}\ell^{2}-4\mu}\,\ell\right)/2}
\end{equation}
and a black hole horizon at
\begin{equation}
R_{\text{bh}}=\sqrt{\left(k\ell^{2}-\sqrt{k^{2}\ell^{2}-4\mu}\,\ell\right)/2}
\end{equation}
for a positive $k$. In this paper we will set $\mu=0$ and $k>0$ for convenience. Therefore, the black hole horizon vanishes and the cosmological horizon reduces to $R_{\text{ch}}=\sqrt{k}\ell$. Note that an observer with velocity $V^{\mu}=(1,0,0,0,0)$ follows a timelike geodesic inside the cosmological horizon.

\begin{figure}[!htb]
\center{
\includegraphics[width=6cm]{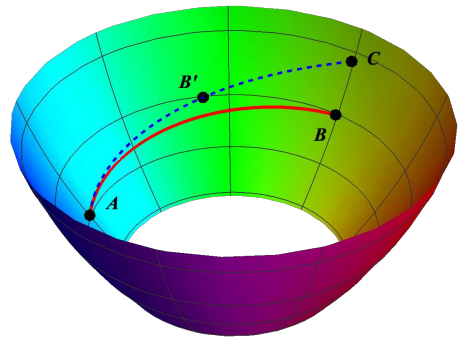}
}
\caption{A schematic picture of our Universe in the $\text{dS}_{5}$ spacetime with a time interval and constants $\theta$ and $\phi$. Each circle on the hypersurface denotes a spacelike curve with constant $R$. Each line orthogonal to the circles on the hypersurface denotes a four-dimensional timelike geodesic evolving with the cosmological time $t$. The blue dashed curve is the trajectory of EMWs on the brane. The red solid curve is the trajectory of GWs in the bulk. Event $A$ denotes the simultaneous emissions of GWs and EMWs. Events $B$ and $C$ mark the detections of the GWs and EMWs on the brane, respectively. $B'$ is the location of the EMWs on the brane when the GWs are detected. It can be seen that there is a time delay for the detection of the EMWs.
}\label{brane}
\end{figure}

To embed our Universe in the $\text{dS}_{5}$ spacetime, we introduce the following constraint
\begin{eqnarray}
-f(R)dT^{2}+f(R)^{-1}dR^{2}=-dt^{2}, \label{c1}
\end{eqnarray}
such that the induced metric of the four-dimensional submanifold (three-dimensional brane and one-dimensional time) coincides with the FLRW metric of our Universe:
\begin{eqnarray}
ds^{2}_{4}=-dt^{2}+R^{2}\Big(\frac{1}{1-kr^{2}}dr^{2}+r^{2}d\theta^{2}+r^{2}\text{sin}^{2}\theta d\phi^{2}\Big)
\end{eqnarray}
with
\begin{eqnarray}
R=a(t)\label{c2}
\end{eqnarray}
on the brane. Here $t$ and $a(t)$ are the cosmological time and the scale factor of our Universe, respectively. The constraint condition~\eqref{c1} implies that the brane is a four-dimensional hypersurface embedded in the $\text{dS}_{5}$ spacetime. The diagram of our Universe with a time interval is shown in Fig.~\ref{brane}. It is obvious that our Universe at each moment corresponds to a three-dimensional sphere $S^3$ with a constant radial distance $R$ on the hypersurface.


In the braneworld scenario, the particles and interactions in the Standard Model are confined on the brane while gravity  propagates through the bulk. Consequently, EMWs and GWs may follow different null curves. Since GWs follow the five-dimensional null geodesics, their trajectory (called shortcut) is expected to be the shortest path from the source to the observer. Assuming that GWs and EMWs are emitted simultaneously, the difference between their trajectories could lead to a time interval between the detections of them and make the viewers on the brane misjudge the speed of GWs. In other words, the four-dimensional viewers may find that the speed of GWs is not equal to the speed of light. In the next section, to figure out this question, we will calculate the gravitational horizon radius and photon horizon radius.

\section{Horizon radius}\label{sec3}

For convenience, we would like to assume that the projections of the events $A$, $B$, $B'$, and $C$ on the 3-sphere (see Fig.~\ref{brane}) are lined along the radial direction $r$. Then the four-dimensional null geodesic connecting events $A$, $B'$, and $C$ and the five-dimensional null geodesic connecting events $A$ and $B$ have the fixed angular variables, $\theta$ and $\phi$. And their trajectories follow
\begin{eqnarray}
ds^{2}_{4}&=&-dt^{2}+a^{2}d\bar{r}^{2},\\
ds^{2}_{5}&=&-f(R)dT^{2}+f(R)^{-1}dR^{2}+R^{2}d\bar{r}^{2},\label{5ng1}
\end{eqnarray}
respectively, where we have introduced a coordinate transformation, i.e., $d\bar{r}^{2}=\frac{1}{1-kr^{2}}dr^{2}$, for the sake of simplification. From the metric~\eqref{5ng1}, one finds two killing vectors, $(\frac{\partial}{\partial T})^{M}$ and $(\frac{\partial}{\partial\bar{r}})^{M}$, and defines the following quantities:
\begin{eqnarray}
\kappa_{T}^{~}&\equiv&g_{MN}^{~}U^{M}\Big(\frac{\partial}{\partial T}\Big)^{N}=-f\frac{dT}{d\lambda},\label{cq1}\\
\kappa_{\bar{r}}^{~}&\equiv&g_{MN}^{~}U^{M}\Big(\frac{\partial}{\partial\bar{r}}\Big)^{N}=R^{2}\frac{d\bar{r}}{d\lambda},\label{cq2}
\end{eqnarray}
where $U^{M}=\frac{dx^{M}}{d\lambda}$ is a unit spacelike vector tangent to the geodesic with $\lambda$ the affine parameter of the five-dimensional null geodesic. Obviously, the quantities $\kappa_{T}^{~}$ and $\kappa_{\bar{r}}^{~}$ are conserved along the five-dimensional null geodesic. For this five-dimensional null geodesic, the combination of Eqs.~\eqref{5ng1},~\eqref{cq1}, and~\eqref{cq2} gives
\begin{eqnarray}
-\frac{\kappa_{T}^{2}}{f}+\frac{1}{f}\Big(\frac{dR}{d\lambda}\Big)^{2}+\frac{\kappa_{\bar{r}}^{2}}{R^{2}}&=&0,\\
\frac{dT}{d\lambda}+\frac{\kappa_{T}^{~}}{f}&=&0,\\
\frac{d\bar{r}}{d\lambda}-\frac{\kappa_{\bar{r}}^{~}}{R^{2}}&=&0,
\end{eqnarray}
and further
\begin{eqnarray}
dR^{2}&=&\frac{R^{2}}{\kappa_{\bar{r}}^{2}}\big(R^{2}\kappa_{T}^{2}-f\kappa_{\bar{r}}^{2}\big)d\bar{r}^{2},\label{i1}\\
dR^{2}&=&\frac{f^{2}}{R^{2}\kappa_{T}^{2}}\big(R^{2}\kappa_{T}^{2}-f\kappa_{\bar{r}}^{2}\big)dT^{2}.\label{i2}
\end{eqnarray}
It is obvious that the comoving distance between events $A$ and $B$ on the brane can be obtained by integrating Eq.~\eqref{i1}:
\begin{eqnarray}
r_{g}&=&\int_{r_{A}}^{r_{B}}\frac{1}{\sqrt{1-kr^{2}}}dr=\int_{\bar{r}_{A}}^{\bar{r}_{B}}d\bar{r}\nonumber\\
&=&\int_{R_{A}}^{R_{B}}\Big(\frac{R^{4}}{s}-R^{2}f\Big)^{-\frac{1}{2}}dR,
\end{eqnarray}
where we have used the definition
\begin{eqnarray}
s\equiv\kappa_{\bar{r}}^{2}/\kappa_{T}^{2}.
\end{eqnarray}
Note that, since the GWs originated from a binary system are finally detected on the brane, the radial distances, $R_{A}$ and $R_{B}$, could be converted into the redshifts of the source and the observer, respectively. Here and after, we rescale the present-day scale factor $a^{~}_{B}$ as the unit and use tildes to denote the rescaled quantities. Accordingly, the rescaled gravitational horizon radius could be expressed as
\begin{eqnarray}
\tilde{r}_{g}\equiv a^{~}_{B}\,r_{g}=\int_{\frac{1}{1+z_{A}}}^{1}\tilde{R}^{-2}\Big(\frac{1}{s}+\frac{1}{\ell^{2}}-\tilde{k}\tilde{R}^{-2}\Big)^{-\frac{1}{2}}d\tilde{R},\label{gh1}
\end{eqnarray}
where $\tilde{k}\equiv k/a_{B}^{2}$ and $\tilde{R}\equiv R/a^{~}_{B}$ are  the rescaled spatial curvature of a 3-sphere and the rescaled radial coordinate, respectively. The observer redshift has been set to zero and the source redshift is marked as $z_{A}$.

Another crucial quantity is the rescaled photon horizon radius:
\begin{eqnarray}
\tilde{r}_{\gamma}\equiv a^{~}_{B}\,r_{\gamma}=\int_{t_{A}}^{t_{B}}\frac{1}{\tilde{a}}dt,\label{eh1}
\end{eqnarray}
where $\tilde{a}\equiv a/a^{~}_{B}$ is the rescaled scale factor. Note that the cosmological times, $t_{A}$ and $t_{B}$, correspond to the moments the GWs are emitted and detected, respectively. As we have mentioned, from the viewpoint of any five-dimensional observer, events $A$ and $B$ are causally connected by a five-dimensional null geodesic. Assuming that the GWs and EMWs are emitted from the binary star merger at the same cosmological time $t_{A}$, the discrepancy between the gravitational horizon radius and photon horizon radius may result in a time delay $\triangle t$ between the detections of the GWs and EMWs. Accordingly, in the case of $\tilde{r}_{g}>\tilde{r}_{\gamma}$, one has an approximate formula
\begin{eqnarray}
\tilde{r}_{g}-\tilde{r}_{\gamma}\approx c\triangle t\label{geh1}
\end{eqnarray}
under the low-redshift case $z\ll0.1$.

Here we should note that if a discrepancy between $\tilde{r}_{g}$ and $\tilde{r}_{\gamma}$ in \eqref{geh1} appears, it will give us an opportunity to explain the detected time delay. In this case, the time delay is expected to carry the information of the extra dimension [see Eqs.~\eqref{gh1} and~\eqref{geh1}] and one could read out the feature of the $\text{dS}_{5}$ radius. Mathematically, if the integral of Eq.~\eqref{gh1} deviates from Eq.~\eqref{eh1}, one can use the observational data of some specific GWs events to constrain the structure of the extra dimension through Eq.~\eqref{geh1}. However, one finds that the quantity $s$ in Eq.~\eqref{gh1} is still unknown. Therefore, one could not obtain the gravitational horizon radius directly from the observed data with Eq.~\eqref{gh1}. To solve this problem, we would use the method introduced in Refs.~\cite{Yu1,Visinelli1} in the next section.

\section{Parameter transformation}\label{sec4}

In the previous section, based on Eq.~\eqref{i1} we have got the gravitational horizon radius on the brane. However, the unobservable quantity $s$ therein makes Eq.~\eqref{gh1} not so practical. To eliminate the quantity $s$, one should recall Eq.~\eqref{i2}, with which the following equation is obtained:
\begin{eqnarray}
\int_{\tilde{T}_{A}}^{\tilde{T}_{B}}d\tilde{T}=\int_{\tilde{R}_{A}}^{\tilde{R}_{B}}\sqrt{\frac{\tilde{R}^{2}}{\tilde{R}^{2}-\tilde{f}s}}\frac{1}{\tilde{f}}d\tilde{R},\label{T1}
\end{eqnarray}
where $\tilde{T}\equiv a^{~}_{B}T$ is the rescaled coordinate time, $\tilde{f}\equiv f/a^{2}_{B}=\tilde{k}-\tilde{R}^{2}/\ell^{2}$ is the rescaled function, and $\tilde{T}^{~}_{A}\equiv a^{~}_{B}T^{~}_{A}$ and $\tilde{T}_{B}^{~}\equiv a^{~}_{B}T_{B}^{~}$ are the rescaled coordinate times of the emission and detection of GWs, respectively. In principle, the quantity $s$ could be solved from this equation directly as long as the coordinate time interval, $\tilde{T}_{AB}\equiv \tilde{T}_{B}-\tilde{T}_{A}$, is known. However, for any four-dimensional observer, the detected time interval is the cosmological time interval, $t_{AB}\equiv t_{B}-t_{A}$, which is related to the source redshift. So one should use Eqs.~\eqref{c1} and~\eqref{c2} to express the rescaled coordinate $\tilde{T}$ in terms of the cosmological time $t$. Accordingly, one finds that the coordinate time interval $\tilde{T}_{AB}$ obeys
\begin{eqnarray}
\int_{\tilde{T}_{A}}^{\tilde{T}_{B}}d\tilde{T}=\int_{t_{A}}^{t_{B}}\frac{\sqrt{F_{1}+H^{2}\tilde{a}^{2}}}{F_{1}}dt,\label{T2}
\end{eqnarray}
where the definition $F_{1}(t)\equiv\tilde{k}+\tilde{a}^{2}/\ell^{2}$ has been employed and the quantity $H(t)$ is the Hubble constant. Based on Eqs.~\eqref{T1} and~\eqref{T2}, the expression of the quantity $s$ in terms of the observable quantities could be obtained by solving the following parameter equation:
\begin{eqnarray}
\int_{\tilde{R}_{A}}^{\tilde{R}_{B}}\sqrt{\frac{\tilde{R}^{2}}{\tilde{R}^{2}-\tilde{f}s}}\frac{1}{\tilde{f}}d\tilde{R}&=&\int_{0}^{z_{A}}\frac{\sqrt{F_{2}+H^{2}}}{F_{2}H}dz,\nonumber\\
&\,&\label{s1}
\end{eqnarray}
where we have utilized the definition $F_{2}(z)\equiv\tilde{k}(1+z)^{2}+\ell^{-2}$. Finally, with Eq.~\eqref{gh1}, one could express the gravitational horizon radius in terms of the $\text{dS}_{5}$ radius and a series of the observable quantities, i.e., the source redshift $z_{A}$, the present-day Hubble constant $H_{B}$, and the present-day spatial curvature density $\Omega_{k}$. Recalling Eq.~\eqref{geh1}, the $\text{dS}_{5}$ radius could be constrained under the low-redshift case for some specific GWs events. In the next section, we will use the data of the event GW170817/GRB 170817A~\cite{Abbott6,Coulter1,Pan1} and the present-day spatial curvature density~\cite{Planck1,Planck2} to constrain the $\text{dS}_{5}$ radius.

\section{Constraint}\label{sec5}

In 2017, the LIGO and Virgo detectors detected the gravitational signal GW170817 originated from the merger of a binary system, which is located at the relatively close distance of $40^{+8}_{-14}\,\text{Mpc}$ from Earth~\cite{Abbott6,Abbott7}. Then,  $1.7\,$s later, GBM and SPI-ACS detected an EM signal, i.e., GRB 170817A, originated from the same place~\cite{Coulter1,Pan1}, which is believed to be an EM counterpart of the event GW170817. The time delay between GW170817 and GRB 170817A opens a wide range of researches on the exotic physics, and could be explained by: $(1)$ the astrophysical process of a binary NS merger~\cite{Shibata1,Rezzolla1,Paschalidis1,Rezzolla2,Ciolfi1,Tsang1}; $(2)$ the modified propagation speed of the GWs~\cite{Comelli1,Will1,Saltas1,Chen1,Rizwana1,Andriot1,Ghosh1}; $(3)$ the shortcut of the five-dimensional null geodesic~\cite{Chung1,Caldwell1,Yu1,Visinelli1,Ishihara1}.

It was found that the hot torus orbiting around a rapidly spinning black hole will be formed within $100\,\text{ms}$ after the binary NS merger and could result in SGRBs~\cite{Shibata1,Rezzolla1,Paschalidis1}. However, the exact process of the binary NS merger is still unknown. So the time-lag $\delta t$ between the emissions of SGRBs and GWs is highly model-dependent. In other words, different astrophysical processes of the binary NS merger predicted in exotic astrophysical models will lead to different time-lags, i.e., $\delta t$'s. In Refs.~\cite{Rezzolla2,Ciolfi1}, the authors pointed out that, if the long-lived binary-merger product forms after the binary NS merger, a black hole-torus system emitting SGRBs will eventually appear through the collapse of the binary-merger product. In such process, SGRBs are expected to be produced later than the peak magnitude of the emission of GWs and $\delta t$ could exceed $10^{3}\,\text{s}$ easily~\cite{Rezzolla2,Ciolfi1}. The time-lag $\delta t$ could be reversed in a different consideration. If the crust-core model is applied to the binary NS merger, then the crust will crack with the emission of GRBs seconds before the merger~\cite{Tsang1}. As a conclusion, these models with different astrophysical processes give a window of the time-lag $\delta t$, i.e., $(-100\,\text{s}, 1000\,\text{s})$~\cite{Abbott8}. Therefore, the astrophysical processes could be used to explain the time delay detected on the observations.

The present four-dimensional gravitational theories could be divided into two categories by Weyl criterion~\cite{Bettoni1}. In the first class of theory, the propagation speed of GWs equals to the speed of light, while in the second class, the speed is not consistent with the speed of light. It is obvious that the time delay $\triangle t$ and the window of the time-lag $\delta t$ of the events GW170817 and GRB 170817A could give a constraint on the second class of theory. According to Ref.~\cite{Abbott8}, by ignoring the contribution of the intergalactic medium dispersion and considering the window of $\delta t$ $(-100\,\text{s}, 1000\,\text{s})$, the speed of the GWs, $c_{g}$, is constrained to
    \begin{eqnarray}
      -2.4\times10^{-13}\leqslant\frac{c_{g}-c_{\gamma}}{c_{\gamma}}\leqslant2.5\times10^{-14}.\label{ccc1}
    \end{eqnarray}
Note that, in Ref.~\cite{Abbott8}, the travel distances of the SGRBs and GWs are chosen as $26$\,Mpc and the window of the time-lag $\delta t$ is (0$\,$s,10$\,$s), so the speed of the GWs, $c_{g}$, is constrained to $-3\times10^{-15}\leqslant\frac{c_{g}-c_{\gamma}}{c_{\gamma}}\leqslant7\times10^{-16}.$
It implies that the modification on the speed of GWs is extremely small.

Note that this constraint is based on the four-dimensional theories. In higher-dimensional theories, EMWs are confined on the brane while GWs could propagate through the bulk. So the trajectory of the GWs could be the shortest path (called shortcut) in the bulk. Therefore, even if the GWs and SGRBs are emitted from the same source simultaneously and follow the null curves, they are expected to arrive at the observer successively and result in the time delay $\triangle t$~\cite{Caldwell1,Yu1,Visinelli1}. In this case, the gravitational horizon radius on the brane is effected by the structure of the extra dimension and the four-dimensional observer may misjudge the propagation speed of the GWs. In conclusion, for a given higher-dimensional gravitational theory which allows a shortcut, the constraint on the propagation speed of the GWs~\eqref{ccc1} will be changed. To show this, we will qualitatively analyse the modified constraint here. As shown in Fig.~\ref{brane}, for the event GW170817/GRB 170817A, when the GWs are detected, the SGRBs is arriving at the point $B'$. Then the timelike curve $BC$ denotes the time delay between the detections of GW170817 and GRB 170817A. For a given higher-dimensional theory and under the low-redshift, the propagation speed of the GWs, $c'_{g}$, viewed by the five-dimensional observer obeys:
\begin{eqnarray}
	\frac{L_{\gamma}}{c_{\gamma}} -\frac{L_{g}}{c'_{g1}} & = &\delta t_{1}+\Delta t\equiv\Delta t_{1},\label{cgu1}\\
	\frac{L_{g}}{c'_{g2}} - \frac{L_{\gamma}}{c_{\gamma}}&= &\delta t_{2}-\Delta t\equiv\Delta t_{2},\label{cgu2}
\end{eqnarray}
where $L_{\gamma}$ is the travel distance of the SGRBs from $A$ to $C$ (see Fig.~\ref{brane}), $L_{g}\equiv L_{\gamma}-\Delta L$ is the travel distance of the GWs from $A$ to $B$ along the shortcut, $\Delta t=1.734\,\text{s}$ is the detected time delay for the event GW170817/GRB 170817A, $c_{\gamma}$ is the propagation speed of the SGRBs on the brane, $\delta t_{1}=100\,\text{s}$ and $\delta t_{2}=1000\,\text{s}$ correspond to the different time-lags between the emissions of the SGRBs and GWs, and $c'_{g1}$ and $c'_{g2}$ are respectively the upper and lower bounds on $c'_{g}$. According to Eqs.~\eqref{cgu1} and~\eqref{cgu2}, the bounds could be expressed as
\begin{eqnarray}
	c'_{g1}&= &\frac{L_{g}c_{\gamma}}{L_{\gamma}-c_{\gamma}\Delta t_{1}}= c^{~}_{g1}-\frac{\Delta L\,c_{\gamma}}{L_{\gamma}-c_{\gamma}\Delta t_{1}},\\
	c'_{g2}&= &\frac{L_{g}c_{\gamma}}{L_{\gamma}+c_{\gamma}\Delta t_{2}}= c^{~}_{g2}-\frac{\Delta L\,c_{\gamma}}{L_{\gamma}+c_{\gamma}\Delta t_{2}},
\end{eqnarray}
or
\begin{eqnarray}
	\frac{c'_{g1}-c_{\gamma}}{c_{\gamma}}&= &\frac{c^{~}_{g1}-c_{\gamma}}{c_{\gamma}}-\delta_{1},\\
	\frac{c'_{g2}-c_{\gamma}}{c_{\gamma}}&= &\frac{c^{~}_{g2}-c_{\gamma}}{c_{\gamma}}-\delta_{2},
\end{eqnarray}
where $\delta_{1}\equiv\frac{\Delta L}{L_{\gamma}-c_{\gamma}\delta t_{1}}$ and $\delta_{2}\equiv\frac{\Delta L}{L_{\gamma}+c_{\gamma}\delta t_{2}}$ are the correction terms determined by the shortcut in the higher-dimensional theory, and $c^{~}_{g1}\equiv\frac{c_{\gamma}L_{\gamma}}{L_{\gamma}-c_{\gamma}\Delta t_{1}}$ and $c^{~}_{g2}\equiv\frac{c_{\gamma}L_{\gamma}}{L_{\gamma}+c_{\gamma}\Delta t_{2}}$ are respectively the upper and lower bounds on $c_{g}$ in the four-dimensional theory. One finds that, for a given higher-dimensional theory allowing a shortcut, the propagation speed of the GWs is modified as follows:
\begin{eqnarray}
	-\big(2.4\times10^{-13}\!+\!\delta_{2}\big)
    \leqslant\frac{c'_{g}\!-\!c_{\gamma}}{c_{\gamma}}
    \leqslant2.5\times10^{-14}\!-\!\delta_{1},
\end{eqnarray}
which shows that both the upper and lower bounds are lower.

In the next context, we will focus on the brane model constructed in Sec.~\ref{sec2} that may allow the existence of the shortcut and assume that GWs and SGRBs are originated from the same source simultaneously with the same speed. Further, we will use the observed time delay $\triangle t$ in the event GW170817/GRB 170817A to constrain the size of the $\text{dS}_{5}$ radius.

\subsection{Approximation}

Before we calculate the constraint on the $\text{dS}_{5}$ radius, we should consider some approximate conditions which are useful to  simplify our calculations in the following context. In Sec.~\ref{sec2}, we have set $k>0$ and $\mu=0$ so that the black hole horizon vanishes. As shown in Fig.~\ref{brane}, our Universe is a four-dimensional hypersurface embedded in the $\text{dS}_{5}$ spacetime. At a constant cosmological time $t_1$, it is a $3$-sphere with a constant radial distance $R_{1}=a(t_{1})$. For big bang theory, our Universe (the brane) arises from a singularity, at which we have $a(t_{0}=0)=0$. With this consideration, our Universe is to be embedded inside the cosmological horizon. On the other hand, one finds that, if events on the brane are causally connected by four-dimensional null or timelike geodesics, they are still causally connected by five-dimensional ones. In other words, the causal structure inside the cosmological horizon is not broken. Note that, as our Universe is inside the cosmological horizon, the present-day radius distance $R_{B}$ of our Universe should be smaller than $\sqrt{k}\ell$, i.e., $\tilde{k}\ell^{2}>1$. Moreover, since the four-dimensional cosmological model will be recovered on the brane in the case of a large $\text{dS}_{5}$ radius~\cite{Verlinde1,Nojiri1,Nojiri2,Addazi1}, we further expect that $R_{B}^{2}\ll k\ell^{2}$. In the following, we will first consider two types of extreme Universe models: de Sitter model (containing only dark energy) and Einstein-de Sitter model (containing only dark matter), which helps us to check the results obtained in the real Universe model. For the real Universe, the results should be somewhere in between. To coincide with the fact that our Universe mainly contains dark energy, dark matter, and ordinary matter, we then assume that the dynamics of the present-day Universe follows the Friedmann equation introduced in the $\Lambda$CDM model. The constraint in this model requires a large $\text{dS}_{5}$ radius.

\subsection{de Sitter model}\label{dSM1}

The four-dimensional Friedmann equation in the de Sitter model is given by
\begin{equation}
    H^{2}=H_{B}^{2}\Big(\Omega_{\Lambda}+\frac{\Omega_{k}}{\tilde{a}^{2}}\Big),\label{dSm}
\end{equation}
where $\Omega_{\Lambda}$ is the present-day effective cosmological constant density and $H_{B}$ is the present-day Hubble constant. Replacing the Hubble constant in Eq.~\eqref{T2} with Eq.~\eqref{dSm}, one finds that the rescaled coordinate time interval $\tilde{T}_{AB}$ between the emission and detection of the GWs obeys the following equation:
\begin{widetext}
\begin{eqnarray} \text{tanh}\Big(\frac{\tilde{T}_{AB}H_{B}\sqrt{-\Omega_{k}}}{\ell}\Big)=\frac{\sqrt{-\Omega_{k}}\sqrt{-1+\ell^{2}H_{B}^{2}(1-\Omega_{k})}(1+z_{A}-\sqrt{1+z_{A}(2+z_{A})\Omega_{k}})}{(1+z_{A})\Omega_{k}\big[1-\ell^{2}H_{B}^{2}(1-\Omega_{k})\big]-\sqrt{1+z_{A}(2+z_{A})\Omega_{k}}}.\label{T3}
\end{eqnarray}
The integral of Eq.~\eqref{T1} has a similar form:
\begin{eqnarray}	
    \text{tanh}\Big(\frac{\tilde{T}_{AB}H_{B}\sqrt{-\Omega_{k}}}{\ell}\Big)=\frac{\sqrt{-\Omega_{k}}\sqrt{-1+Q\ell^{2}}\Big((1+z_{A})\sqrt{\frac{Q}{H_{B}^{2}}+\Omega_{k}}-\sqrt{\frac{Q}{H_{B}^{2}}+(1+z_{A})^{2}\Omega_{k}}\Big)}{(1+z_{A})\Omega_{k}(1-Q\ell^{2})-\sqrt{\frac{Q}{H_{B}^{2}}+\Omega_{k}}\sqrt{\frac{Q}{H_{B}^{2}}+(1+z_{A})^{2}\Omega_{k}}},\label{T4}
\end{eqnarray}
\end{widetext}
where $Q\equiv\frac{1}{s}+\frac{1}{\ell^{2}}$ is a new parameter introduced for convenience. It is obvious that Eqs.~\eqref{T3} and~\eqref{T4} are equivalent as long as
$s=\ell^{2}[H_{B}^{2}(1-\Omega_{k})\ell^{2}-1]^{-1}$. So it is the solution of Eq.~\eqref{s1}. Finally, one obtains the gravitational horizon radius in terms of the observable quantities as follows:
\begin{eqnarray}
	\tilde{r}_{g}=\frac{\text{arctan}\bigg(\frac{\sqrt{-\Omega_{k}}}{\sqrt{\Omega_{k}+\frac{1-\Omega_{k}}{(1+z_{A})^{2}}}}\bigg)-\text{arctan}(\sqrt{-\Omega_{k}})}{H_{B}\sqrt{-\Omega_{k}}},
\end{eqnarray}
which indeed equals to the photon horizon radius~\eqref{eh1}. It implies that, in our model, if the Universe is dominated by the dark energy, the extra dimension may not cause the discrepancy between the gravitational horizon radius and photon horizon radius. Therefore, there is no time delay $\triangle t$ between the GWs and EMWs emitted simultaneously.

One should note that Eq.~\eqref{T2} highly depends on cosmological models. So one may get a different solution of the quantity $s$ in other cosmological models. In the next subsection, we will give the gravitational and photon horizon radii in the Einstein-de Sitter model.

\subsection{Einstein-de Sitter model}\label{EdSm}

Now, we assume that our Universe is dominated by the dark matter. In this case Eq.~\eqref{T3} and hence the solution of $s$ will change. Since the calculation is extremely tedious, we do not show all results here. Instead, for the event GW170817/GRB 170817A, we give the gravitational horizon radius and photon horizon radius as follows:
\begin{eqnarray}
\tilde{r}_{g}&\simeq&\frac{\text{arctan}\bigg(\frac{H_{B}\sqrt{-\Omega_{k}}}{\sqrt{\frac{C_{2}}{(1+z_{A})^{2}}+H_{B}^{2}\Omega_{k}}}\bigg)-\text{arctan}\Big(\frac{H_{B}\sqrt{-\Omega_{k}}}{\sqrt{C_{2}+H_{B}^{2}\Omega_{k}}}\Big)}{H_{B}\sqrt{-\Omega_{k}}}\nonumber\\
&~&+\frac{\sqrt{C_{2}+H_{B}^{2}\Omega_{k}}-\sqrt{\frac{C_{2}}{(1+z_{A})^{2}}+H_{B}^{2}\Omega_{k}}}{2\sqrt{\frac{C_{2}}{(1+z_{A})^{2}}+H_{B}^{2}\Omega_{k}}\sqrt{C_{2}+H_{B}^{2}\Omega_{k}}}\frac{C_{1}}{C_{2}\ell^{2}H_{B}^{2}\Omega_{k}},\nonumber\\
&~&\label{gh2}\\
\tilde{r}_{\gamma}&=&\frac{2\Bigg[\text{arctan}(\sqrt{-\Omega_{k}})-\text{arctan}\bigg(\frac{\sqrt{-\Omega_{k}}}{\sqrt{1+z_{A}-z_{A}\Omega_{k}}}\bigg)\Bigg]}{H_{B}\sqrt{-\Omega_{k}}}.\label{eh2}
\end{eqnarray}
Here $C_{1}$ and $C_{2}$ are very complex parameter functions of $z_{A}$, $H_{B}$, and $\Omega_{k}$. We also do not show them here. Generally speaking, the two horizon radii~\eqref{gh2} and~\eqref{eh2} are different and the discrepancy between them will result in the time delay $\triangle t$. Taking advantage of the events GW170817 and GRB 170817A, we can obtain the constraint on the $\text{dS}_{5}$ radius.

\begin{figure*}[!htb]
\center{
\subfigure[]{\includegraphics[width=5cm]{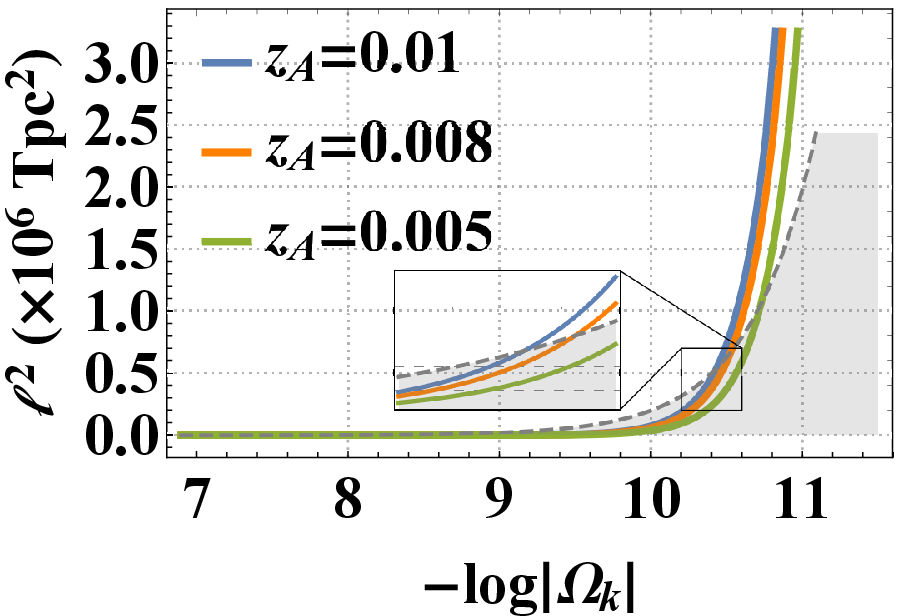}\label{constraint1}}
\quad
\subfigure[]{\includegraphics[width=5cm]{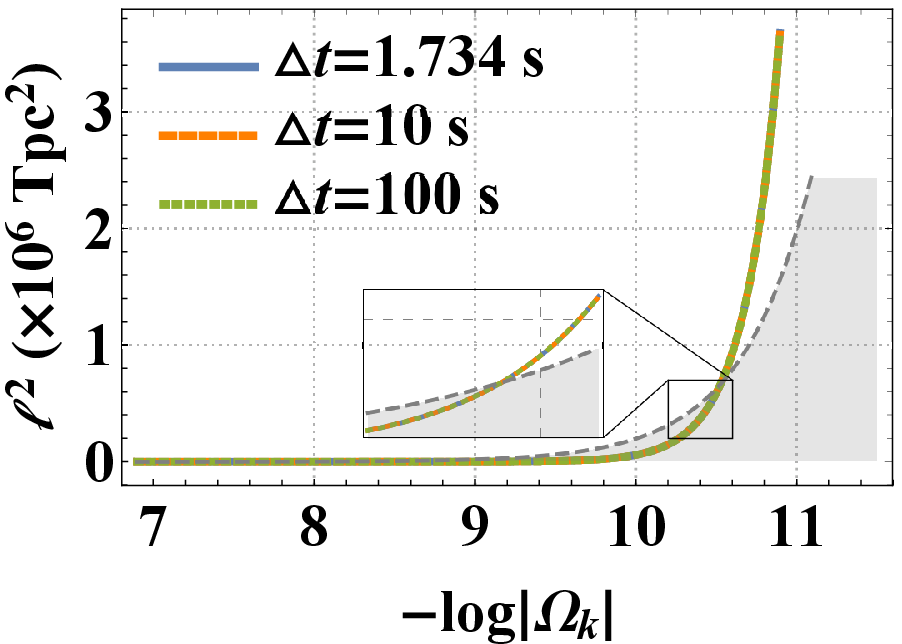}\label{constraint2}}
\quad
\subfigure[]{\includegraphics[width=5cm]{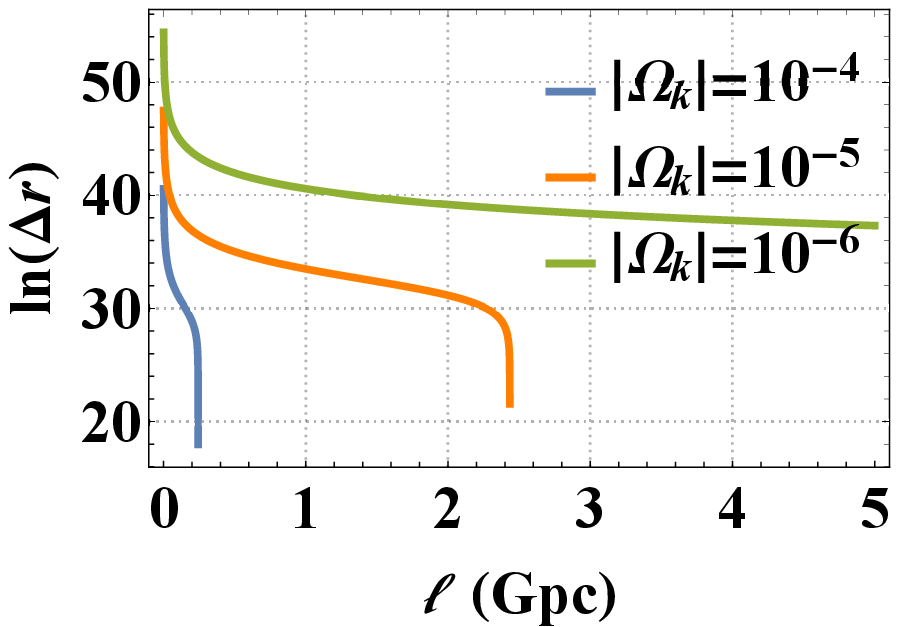}\label{constraint3}}
}
\caption{The $\text{dS}_{5}$ radius with respective to the present-day spatial curvature density and the discrepancy between the gravitational and photon horizon radii  with respect to the $\text{dS}_{5}$ radius for the event GW170817/GRB 170817A. (a) The $\text{dS}_{5}$ radius for different cosmological redshifts. The redshifts, $z_{A}=0.005$ (green curve), $z_{A}=0.008$ (orange curve), and $z_{A}=0.01$ (blue curve), correspond to the data given by the gravitational observation of GW170817 with the consideration of the uncertainty. The time delay $\triangle t$ is fixed to $\Delta t=1.734\,\text{s}$. (b) The $\text{dS}_{5}$ radius for different time delays $\triangle t$'s. We fix the redshift to $z_{A}=0.008$ in this subfigure. The blue curve with $\Delta t=1.734\,\text{s}$ corresponds to the assumption that GWs and SGRBs are originated from the source simultaneously. According to the window $(-100\,\text{s},\,1.734\,\text{s})$ predicted in the exotic astrophysical models, the curve, $\Delta t=100\,\text{s}$ (green dotted curve), denotes the case that SGRBs are emitted from the binary NS merger about $100\,\text{s}$ early than the emission of GWs. The orange dashed curve with $\Delta t=10\,\text{s}$ is used to show the $\text{dS}_{5}$ radius in the case of a time-lag $\delta t$ within ($-100$\,s, $1.734$\,s). The gray dashed region in the first two subfigures denotes the region excluded by the condition $\tilde{k}\ell^{2}>1$. (c) The dimensionless discrepancy $\Delta r\equiv(\tilde{r}_{g}-\tilde{r}_{\gamma})/L$ with $L=1\,\text{Mpc}$ for different present-day spatial curvature densities. The blue, orange, and green curves correspond to the different present-day spatial curvature densities: $|\Omega_{k}|=10^{-4}$, $|\Omega_{k}|=10^{-5}$, and $|\Omega_{k}|=10^{-6}$.}
\label{constraint}
\end{figure*}

From the data of the LIGO-Virgo detectors, the source redshift of  GW170817 is about $z_{A}=0.008^{+0.002}_{-0.003}$~\cite{Abbott6,Abbott7}. The collaboration of the LIGO-Virgo detectors, GBM, and SPI-ACS shows a time delay about $\triangle t=1.74^{+0.05}_{-0.05}\,\text{s}$ between the detections of GW170817 and its EM counterpart GRB 170817A~\cite{Coulter1,Pan1,Abbott6,Abbott7,Abbott8}. Concerning the uncertainty, we choose three sets of the redshifts ($0.01$, $0.008$, $0.005$). In addition, we first assume that the GWs and EMWs are emitted simultaneously. Then, with respect to the present-day spatial curvature density, the $\text{dS}_{5}$ radius is shown in Fig.~\ref{constraint1}. Based on the {\em Planck} data~\cite{Planck1,Planck2}, the present-day Hubble constant is chosen as $H_{B}=67.36\,\text{km}\,\text{s}^{-1}\,\text{Mpc}$.

From Fig.~\ref{constraint1}, one finds that the $\text{dS}_{5}$ radius increases with the decrease of $|\Omega_{k}|$. The uncertainty in the measured source redshift of GW170817 could result in a small correction to the constraint on the $\text{dS}_{5}$ radius. It means that the $\text{dS}_{5}$ radius is not sensitive to the redshift ranging from $0.005$ to $0.01$. The analysis of the cosmic microwave background by {\em Planck} collaboration has given a constraint on the present-day spatial curvature density $|\Omega_{k}|\lesssim10^{-3}$ (see Table~$5$ in Ref.~\cite{Planck1} and Table~$4$ in Ref.~\cite{Planck2}). According to it, we constrain the $\text{dS}_{5}$ radius to $\ell\gtrsim 27.1\,\text{Mpc}$ with the combination of Eqs.~\eqref{geh1}, \eqref{gh2}, and~\eqref{eh2}. On the other hand, since the embedded Universe requires an extra condition on the $\text{dS}_{5}$ radius, i.e., $\tilde{k}\ell^{2}>1$, the $\text{dS}_{5}$ radius located in the gray dashed region is excluded out. Consequently, it gives some stronger constraints on the $\text{dS}_{5}$ radius, e.g.,  $\ell\gtrsim7.5\times10^{2}\,\text{Tpc}$ for $z_{A}=0.01$ and $\ell\gtrsim1.1\times10^{3}\,\text{Tpc}$ for $z_{A}=0.005$. Accordingly, we finally give a lower bound on the $\text{dS}_{5}$ radius $\ell\gtrsim7.5\times10^{2}\,\text{Tpc}$.
As shown in Fig.~\ref{constraint1}, the bound also denotes a constraint on the present-day spatial curvature density $|\Omega_{k}|\lesssim3.6\times10^{-11}$. Therefore, if the constraint on the spatial curvature density obtained by the future {\em Planck} collaboration contradicts to this result, the five-dimensional model considered in this paper should be ruled out then.

The previous context mainly bases on the assumption that the GWs and SGRBs are emitted simultaneously. Here we will consider the contribution of different astrophysical processes of the binary NS merger to the $\text{dS}_{5}$ radius. Since we have assumed that the GWs have the same speed as the SGRBs, the shortcut will naturally result in a larger gravitational horizon radius than the photon horizon radius. Therefore, we choose the window ranging from $-100\,\text{s}$ to $1.734\,\text{s}$ and investigate the effect of them on the $\text{dS}_{5}$ radius. As shown in Fig.~\ref{constraint2}, an interesting result is that such time-lag $\delta t$ introduced in the present alternative astrophysical models has little effect on the constraint. It could be explained by the fact that the dimensionless discrepancy $\Delta r\equiv(\tilde{r}_{g}-\tilde{r}_{\gamma})/L$ with $L=1\,\text{Mpc}$  between the two horizon radii decreases extremely fast with the increase of the $\text{dS}_{5}$ radius [see Fig.~\ref{constraint3}].

\subsection{$\Lambda$CDM model}

The {\em Planck} results~\cite{Planck1,Planck2,Planck3} indicate that our Universe could be well described by a spatially flat $\Lambda$CDM model. Therefore, we will assume the expansion rate of our real Universe as
\begin{equation}
	H^{2}=H_{B}^{2}\Big(\Omega_{\Lambda}+\frac{\Omega_{k}}{\tilde{a}^{2}}+\frac{\Omega_{m}}{\tilde{a}^{3}}\Big),\label{lambdacdmm}
\end{equation}
where $\Omega_{m}$ is the present-day energy density of the nonrelativistic matter, $\Omega_{\Lambda}$ is the present-day energy density of the dark energy, and $\Omega_{k}$ is small compared with $\Omega_{m}$ and $\Omega_{\Lambda}$. With the same approach used in the previous subsections, we obtain the gravitational horizon for the event GW170817/GRB 170817A:
\begin{eqnarray}\label{gh3}
	\tilde{r}_{g}&\simeq&\frac{\text{arctan}\bigg(\frac{H_{B}\sqrt{-\Omega_{k}}}{\sqrt{\frac{C_{3}}{(1+z_{A})^{2}}+H_{B}^{2}\Omega_{k}}}\bigg)-\text{arctan}\bigg(\frac{H_{B}\sqrt{-\Omega_{k}}}{\sqrt{C_{3}+H_{B}^{2}\Omega_{k}}}\bigg)}{H_{B}\sqrt{-\Omega_{k}}}\nonumber\\
	&~&+\frac{\sqrt{C_{3}+H_{B}^{2}\Omega_{k}}-\sqrt{\frac{C_{3}}{(1+z_{A})^{2}}+H_{B}^{2}\Omega_{k}}}{2\sqrt{C_{3}+H_{B}^{2}\Omega_{k}}\sqrt{\frac{C_{3}}{(1+z_{A})^{2}}+H_{B}^{2}\Omega_{k}}}\frac{C_{4}}{C_{3}\ell^{2}H^{2}_{B}\Omega_{k}},\nonumber\\
\end{eqnarray}
which has the same form as Eq.~\eqref{gh2} while $C_{3}$ and $C_{4}$ are parameter functions of $\Omega_{\Lambda}$, $\Omega_{m}$, $\Omega_{k}$, $H_{B}$, and $z_{A}$. Obviously, $C_{3}$ and $C_{4}$ are much more complex than $C_{1}$ and $C_{2}$ and we are not going to show the explicit forms of them. From the analysis in the previous subsections, we know that the time delay $\triangle t$ resulting from the shortcut would arise when the expansion rate of our Universe deviates from Eq.~\eqref{dSm}. Therefore, for the same value of $\Omega_{k}$, such time delay $\triangle t$ should be smaller in the $\Lambda$CDM model compared with the Einstein-de Sitter model. On the other hand, the similar form of Eqs.~\eqref{gh2} and~\eqref{gh3} indicates that $\triangle t\propto1/\ell^{2}$. One then expects that the lower bound on the $\text{dS}_{5}$ radius in $\Lambda$CDM model is larger than the one in the Einstein-de Sitter model.

\begin{figure*}[!htb]
\center{
\subfigure[]{\includegraphics[width=5cm]{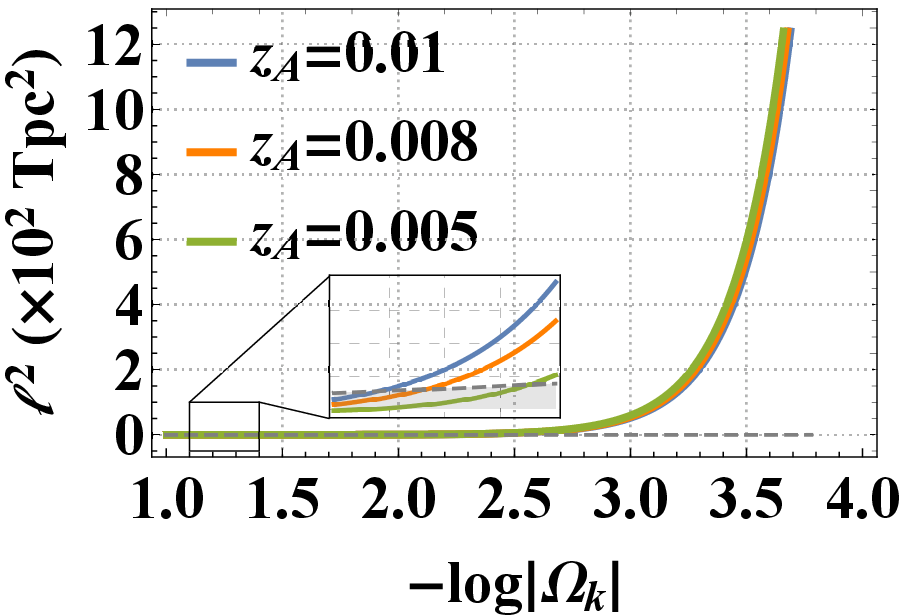}\label{constraint4}}
\quad
\subfigure[]{\includegraphics[width=5cm]{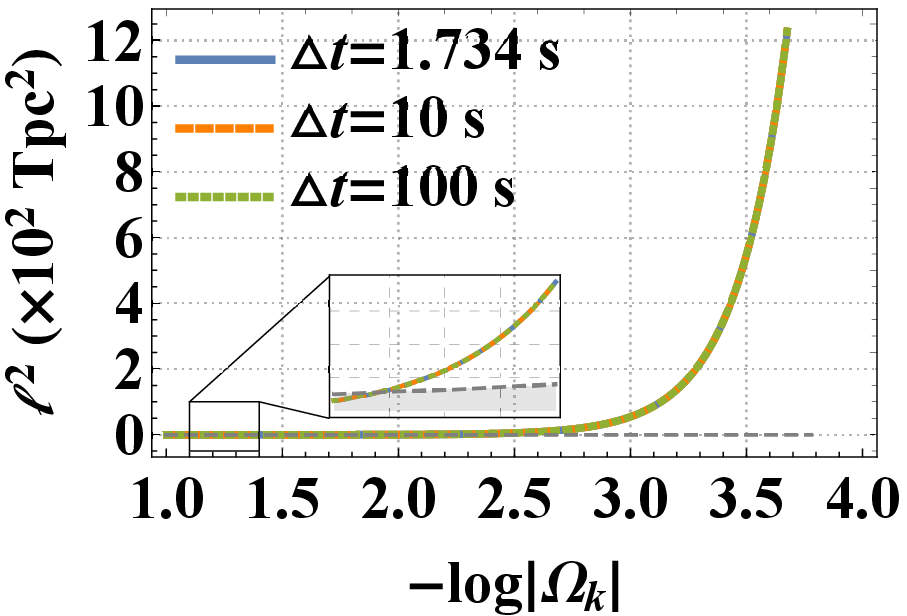}\label{constraint5}}
}
\caption{The $\text{dS}_{5}$ radius with respective to the present-day spatial curvature density. (a) The $\text{dS}_{5}$ radius for different cosmological redshifts. The redshifts and the time delay $\triangle t$ are fixed to $z_{A}=0.005$ (green curve), $z_{A}=0.008$ (orange curve), $z_{A}=0.01$ (blue curve), and $\Delta t=1.734\,\text{s}$, respectively. (b) The $\text{dS}_{5}$ radius for different time delays $\triangle t$'s. We fix the redshift to $z_{A}=0.008$ in this subfigure. We choose three kinds of curves, the blue curve with $\Delta t=1.734\,\text{s}$, the orange dashed curve with $\Delta t=10\,\text{s}$, and the green dotted curve with $\Delta t=100\,\text{s}$, according to the window of the time-lag $\delta t$, $(-100\,\text{s},\,1.734\,\text{s})$, predicted in the exotic astrophysical models. The gray dashed region in the first two subfigures denotes the region excluded by the condition $\tilde{k}\ell^{2}>1$.}
\label{constraintp2}
\end{figure*}

In the following analysis, we will fix the density parameters to $\Omega_{\Lambda}=0.685$ and $\Omega_{m}=0.315$, according to the {\em Planck} 2015 and {\em Planck} 2018 data~\cite{Planck1,Planck2}. Besides, we will respectively parametrize $\Omega_{\Lambda}$ and $\Omega_{m}$ as
\begin{equation}
	\Omega_{\Lambda}=\frac{\Omega_{\Lambda}}{\Omega_{\Lambda}+\Omega_{m}}(1-\Omega_{k})
\end{equation}
and
\begin{equation}
	\Omega_{m}=\frac{\Omega_{m}}{\Omega_{\Lambda}+\Omega_{m}}(1-\Omega_{k}),
\end{equation}
such that the ratio between the dark energy and the nonrelativistic matter is unchanged when considering the contribution of the curvature. Taking advantage of the events GW170817 and GRB 170817A, we show the constraints on the $\text{dS}_{5}$ radius in Fig.~\ref{constraintp2}. Similar to the case of the Einstein-de Sitter model, the uncertainty in the redshift and the different time-lags $\delta t$s have little effect on the constraint. While, for the same value of the present-day curvature energy density, $\Lambda$CDM model requires a larger $\text{dS}_{5}$ radius compared with the Einstein-de Sitter model such that the condition $\tilde{k}\ell^{2}>1$ (gray dashed region) gives a smaller lower bound on the $\text{dS}_{5}$ radius, e.g., $\ell\gtrsim16.5\,\text{kpc}$ for $z_{A}=0.01$ and $\ell\gtrsim19.3\,\text{kpc}$ for $z_{A}=0.005$. From Fig.~\ref{constraintp2}, one finds that these constraints are much smaller than the constraint given by the condition $|\Omega_{k}|\lesssim10^{-3}$. Eventually, with $|\Omega_{k}|\lesssim10^{-3}$, we give a lower bound on the $\text{dS}_{5}$ radius as $\ell\gtrsim2.4\times10^{3}\,\text{Tpc}$ which is larger than the lower bound in the Einstein-de Sitter model (see Sec.~\ref{EdSm}). It is now obvious that, in $\Lambda$CDM model, a large $\text{dS}_{5}$ radius is required, and this result is quite different from the case in Ref.~\cite{Visinelli1} (where $\text{AdS}_{5}$ radius is constrained to $\ell\gtrsim0.535\,\text{Mpc}$). As we have mentioned before, the {\em Planck} analyses indicate that the standard $\Lambda$CDM model could describe our real Universe in a high accuracy~\cite{Planck1,Planck2,Planck3}. Therefore, the modified Friedmann equation on the brane should not deviate from the standard form too much. For an $\text{AdS}_{5}$ bulk, to ensure that the modified field equation on the brane recovers the Einstein-Hilbert equation, one needs a small $\text{AdS}_{5}$ radius~\cite{ArkaniHamed1998rs,Randall1,Randall2}. While, for a $\text{dS}_{5}$ bulk, as shown in Refs.~\cite{Verlinde1,Nojiri1,Nojiri2,Addazi1}, one recovers the standard cosmology model only if the $\text{dS}_{5}$ radius is extremely large. It is then obvious that our result coincides with this fact.

\section{Conclusion}\label{sec6}

The recent detection of GW170817 by the LIGO-Virgo detectors and the following detection of GRB 170817A by GBM and SPI-ACS imply a binary NS merger near the NGC 4399~\cite{Abbott6,Coulter1,Pan1} and prove the link between SGRBs and binary NS mergers~\cite{Abbott7}. It therefore opens a wide range of astrophysical researches on the NS (e.g. Refs.~\cite{Abbott9,Margalit1,Pian1,Drout1,Annala1,Most1,Fattoyev1}). On the other hand, the observed time delay $\triangle t$ between the GWs and SGRBs may reveal the existence of exotic physics. In higher-dimensional theories, GWs could pass through the bulk while EMWs are confined on the brane (see Fig.~\ref{brane}). Assuming that both the GWs and SGRBs follow null curves and are originated from the same source simultaneously, a discrepancy between the gravitational horizon radius and photon horizon radius may appear and result in the time delay observed by the detectors. Since the gravitational horizon radius is affected by the structure of the extra dimension, it is then expected that the time delay will carry some information of the extra dimension. Therefore, the time delay is a key to investigating the structure of the extra dimension~\cite{Caldwell1,Yu1,Visinelli1}.

We considered a static spherically symmetric $\text{dS}_{5}$ spacetime, where our Universe (viewed as a brane) is embedded. The brane is located inside the cosmological horizon so that the scale factor could increase from $a(t=0)=0$ to $a(t=t_B)$. We firstly calculated the gravitational horizon radius and photon horizon radius under the assumption that our Universe is dominated by the dark energy/the dark matter. We found that, in the former case, the discrepancy between the two horizon radii vanishes, while in the latter case, it appears and the observed time delay may result from it. Accordingly, we used the time delay $\triangle t$ to constrain the $\text{dS}_{5}$ radius with respect to different present-day spatial curvature densities (see Fig.~\ref{constraint}). We found that the constraint on $\Omega_{k}$ given by the {\em Planck} collaboration~\cite{Planck1,Planck2} leads to a constraint on the $\text{dS}_{5}$ radius $\ell\gtrsim 27.1\,\text{Mpc}$. Moreover, the extra condition, $\tilde{k}\ell^{2}>1$, used to embed our Universe excludes out some parameter regions and gives a stronger constraint on the $\text{dS}_{5}$ radius, e.g., $\ell\gtrsim7.5\times10^{2}\,\text{Tpc}$ for $z_{A}=0.01$ and $\ell\gtrsim1.1\times10^{3}\,\text{Tpc}$ for $z_{A}=0.005$. Consequently, we gave a lower bound on the $\text{dS}_{5}$ radius as $\ell\gtrsim7.5\times10^{2}\,\text{Tpc}$ in the Einstein-de Sitter model. We also found that this bound gives a constraint on the present-day spatial curvature density $|\Omega_{k}|\lesssim3.63\times10^{-11}$, which is useful to judge the validity of the five-dimensional model with an Einstein-de Sitter brane.

To coincide with the observations~\cite{Planck1,Planck2,Planck3}, we then consider the $\Lambda$CDM model of our Universe where the dark energy and the dark matter are both included. Unlike the case of the Einstein-de Sitter model, the condition $\tilde{k}\ell^{2}>1$ now constrains the $\text{dS}_{5}$ radius to, e.g., $\ell\gtrsim16.5\,\text{kpc}$ for $z_{A}=0.01$ and $\ell\gtrsim19.3\,\text{kpc}$ for $z_{A}=0.005$. Besides, as shown in Fig.~\ref{constraintp2}, the constraint $|\Omega_{k}|\lesssim10^{-3}$ gives a smaller lower bound as $\ell\gtrsim2.4\times10^{3}\,\text{Tpc}$ compared with the Einstein-de Sitter model. Indeed, by comparing Eqs.~\eqref{gh2} and~\eqref{gh3}, one finds that the time delay $\Delta t$ obeys $\Delta t\propto1/\ell^{2}$. Note that, in Sec.~\ref{dSM1}, we have pointed out that the time delay resulting from the extra dimension vanishes in the de Sitter model. Though no constraint could be obtained directly in this model, we could give an estimation on the $\text{dS}_{5}$ radius $\ell$ by taking the condition $\Omega_{\Lambda}\gg\Omega_{m}$ in the $\Lambda$CDM model. Since the time delay is proportional to the inverse of $\ell^{2}$, a vanishing contribution of the extra dimension to the time delay denotes an extremely large $\text{dS}_{5}$ radius in the de Sitter model. On the other hand, by taking into account the condition $\Omega_{m}\gg\Omega_{\Lambda}$, the constraint obtained in the $\Lambda$CDM model approximates to the one in the Einstein-de Sitter model. Thus the de Sitter model gives an upper bound on the $\text{dS}_{5}$ radius, while the Einstein-de Sitter model gives a lower bound. The constraint obtained in the $\Lambda$CDM model should be somewhere in between. Our results are consistent with it.

We also simply analyze the sensitivity of our constraint to the parameters. We investigated the effect of the uncertainty in the source redshift on our constraint. The result shows that, for the event GW170817, the uncertainty gives little effect on the constraint. Note that this constraint was obtained by assuming the simultaneous emissions of the GWs and SGRBs. However, for different astrophysical processes of the binary NS merger, there is a time-lag between the emissions of the GWs and SGRBs. Therefore, we chose some nonvanishing $\delta t$'s to show the effect of different astrophysical processes on the constraint. Since the GWs follow the shortcut which is a null curve, we considered a window of the time-lag $\delta t$ ranging from $-100\,\text{s}$ to $1.734\,\text{s}$. We finally found that, our constraints are not sensitive to the time-lag in this range [see Figs.~\ref{constraint2} and~\ref{constraint5}].

\section*{Acknowledgements}

We thank Si-Jiang Yang, Hai Yu, and Xiang-Ru Li for useful discussions. We also thank the referee for his/her helpful suggestions.
This work was supported by the National Natural Science Foundation of China (Grants No.~11875151 and No.~11522541), the Strategic Priority Research Program on Space Science, the Chinese Academy of Sciences (Grant No.~XDA15020701), and the Fundamental Research Funds for the Central Universities (Grants No.~lzujbky-2020-it04 and No.~lzujbky-2019-ct06).

\end{document}